\documentstyle[epsfig,prl,twocolumn,aps,floats]{revtex}

\begin{document}
\topmargin = -2.0cm
\overfullrule 0pt
\twocolumn[\hsize\textwidth\columnwidth\hsize\csname
@twocolumnfalse\endcsname


\title{Decay of accelerated protons and the existence of the
Fulling-Davies-Unruh effect}


\author{Daniel A. T. Vanzella and George E. A. Matsas}
\address{Instituto de F\'\i sica Te\'orica, Universidade Estadual Paulista,
         Rua Pamplona 145, 01405-900, S\~ao Paulo, S\~ao Paulo,
         Brazil}

\maketitle
\vspace{.5cm}
\hfuzz=25pt
\draft

\begin{abstract}
We investigate the weak decay of uniformly {\em accelerated protons} in the
context of {\em standard} Quantum Field Theory.  Because the mean {\em proper}
lifetime of a particle is a scalar, the same value for this observable must be
obtained in the inertial and coaccelerated frames.  We are only able to achieve
this equality by considering the Fulling-Davies-Unruh effect.  This reflects 
the fact that the Fulling-Davies-Unruh effect is mandatory for the 
consistency of Quantum Field Theory.  There is no question about its 
existence provided one accepts the validity of standard Quantum Field
Theory in flat spacetime.
\end{abstract}
\vskip2pc]

A couple of years after the discovery by Hawking that black holes should
evaporate~{\cite{H}}, Unruh realized that many features present in the Hawking
effect could be better understood in the simpler context of Minkowski
spacetime~\cite{U}.  As an extra bonus, he found that the Minkowski vacuum, 
i.e., the quantum state associated with the nonexistence of particles according
to inertial observers, corresponds to a thermal bath of elementary particles 
at temperature $T_{\rm FDU}= a \hbar / 2 \pi k c$ as measured by uniformly
accelerated observers with proper acceleration $a$.  Indeed this reflects the
fact that the particle content of a Quantum Field Theory (QFT) is observer 
dependent, as noted by Fulling~\cite{F} and Davies~\cite{D} some time before.
Thus while inertial observers in Minkowski vacuum would be frozen at 
$0 \; {\rm K}$, accelerated ones would be burnt provided that their proper 
acceleration were high enough.

Perhaps partly because of its ``paradoxical-looking'' and partly because of 
the technicalities involved in its derivation (see, e.g., 
Ref.~\cite{Waldbook}),  the Fulling-Davies-Unruh (FDU) effect is still source 
of much skepticism. As a consequence, 
much effort has been spent to devise ways of observing it 
(see, e.g., Ref.~\cite{R} and references therein for a comprehensive list).
Since $T_{\rm FDU}= [a/(2.5 \times 10^{22}\; {\rm cm/s}^2)]$~K, direct 
manifestations of the FDU effect would only be expected under extremely 
high acceleration regimes. Very recently, e.g., Chen and Tajima suggested the 
possibility of observing the FDU effect by means of Petawatt-class lasers 
with which $e^-$'s would reach accelerations of 
$\sim 10^{28} {\rm cm/s}^2$ in every laser cycle~\cite{CT}. 
It is well known that accelerated $e^-$'s  suffer recoil because 
of the radiation reaction force associated with the Larmor radiation.
{\em For instance,} an $e^-$  in a constant electric field ${\bf E}$ should 
quiver around  a uniformly accelerated worldline with 
proper acceleration $a=e |{\bf E}| /m_e$, where $e$ and $m_e$ are 
the electron charge and mass, respectively. Rather than using the radiation 
reaction  force to calculate the $e^-$ recoil, Chen and Tajima have estimated 
it by assuming that the quivering is a consequence of the random  absorption 
of quanta from the FDU thermal bath as seen in the $e^-$'s proper frame. 
Inspired by this, they call the  recoil-induced photon emission ``Unruh 
radiation''. Eventually they calculate the emitted power associated with 
the Unruh radiation for an $e^-$ during each laser half-cycle 
and argue that its observation would consist of an experimental test for 
the FDU effect.

Here we would like to look at this issue from a distinct point of view.  Rather
than looking for an experimental manifestation of the FDU effect when high
accelerations are achieved, which, in general, leads to paramount technical
problems~\cite{L}, we will take a theoretic-oriented strategy.  This sort of
approach is not new~\cite{UW}-\cite{GM} but we hope that the comprehensive
understanding  brought by the FDU effect to the decay of 
{\em accelerated} $p^+$'s 
(which is a potentially important phenomenon in its own right) 
will be very convincing of
the necessity of this effect for the consistency of QFT.  
First, we will  analyze in the inertial frame and
using {\em standard} QFT the  decay of uniformly accelerated
$p^+$'s and next we will show that the FDU effect is essential to reproduce
the {\em proper} decay rate in the uniformly accelerated frame.

According to the Standard Model, {\em inertial} $p^+$'s are stable, 
which is in agreement with highly accurate experiments  ($\tau_{p } > 1.6 
\times 10^{25}$ years)~\cite{PDG}.  As far as we know, the first ones to 
comment that noninertial $p^+$'s could decay were Ginzburg and 
Syrovatskii~\cite{GS} but no calculations were performed until Muller~\cite{M}
obtained an estimation of the decay rate associated with the process 
$$
{\rm (i)}\;\;\;\; p^+ \stackrel{a}{\to} n^0\; e^+ \nu_e \;\;\; 
$$
by assuming that all the involved particles are scalars. 
A more realistic calculation describing the leptons as fermions was only 
performed very recently by the authors~\cite{VM}. The  
energy scale of the emitted particles in the $p^+$ instantaneous inertial rest 
frame  is of order of the $p^+$ proper acceleration $a$.  Thus if 
$a \ll m_{Z^0}, m_{W^\pm}\; (\approx 10^{36}{\rm cm/s}^2)$,
a Fermi-like effective theory can be used. The effective coupling 
constant is fixed such that the $\beta$-decay rate for {\em inertial} 
$n^0$'s  be compatible with observation, i.e., leads to a
mean proper lifetime of $887 \;{\rm s}$~\cite{PDG}.

Protons are not likely to decay in laboratory conditions, e.g,
at LHC/CERN $a \approx 10^{23}\;{\rm cm/s^2}$ in which case
the $p^+$ mean lifetime is
$
\tau_p \approx 10^{3\times 10^8}{\rm yr} \; .
$
Notwithstanding  some astrophysical situations are much more promising.
A cosmic ray $p^+$ with energy $E_p  \approx 1.6 \times 10^{14}$ eV 
under the influence of the magnetic field $B \approx 10^{14}$ G
of a typical pulsar has a proper acceleration  
$a \approx 5 \times 10^{33} {\rm cm/s^2}$ and is confined in a 
cylinder of radius  $R \approx  5\,\times \, 10^{-3} \; {\rm cm}
\ll l_B , $ where $l_B$ is the typical size of the magnetic field region.
As a result, $p^+$'s would have a mean ``laboratory'' lifetime of
$t_p \approx 10^{-1}$ s. For $l_B \approx 10^7 \; {\rm cm}$,
we  obtain that about $| \Delta N_p /N_p | =(1-e^{-l_B/t_p}) \approx l_B/t_p 
\approx 1\%$ of the $p^+$'s would decay via reaction~(i). 
For a potentially interesting relation between the {\em strong} decay of 
accelerated $p^+$'s and the central engines of gamma-ray bursts
obtained with the idealization that $p^+$'s and $n^0$'s 
have the same mass, see Ref.~\cite{TK}.

For our present purposes it is enough to analyze reaction (i) in a 
2-dimensional spacetime. Hereafter we  use signature
$(+ -)$ and natural  units $k_{B} = c = \hbar = 1$ unless stated 
otherwise. The worldline of a uniformly accelerated $p^+$ in usual 
Cartesian coordinates of Minkowski spacetime is given by 
$z^2 - t^2 = a^{-1}$ where $\sqrt{a^\mu a_\mu}= a = {\rm const}$ 
is the $p^+$ proper acceleration. We construct, thus, the vector 
current $j^\mu = q u^\mu \delta (\sqrt{z^2-t^2\,} - a^{-1}) $
associated with a uniformly accelerated classical $p^+$ with 4-velocity 
$u^\mu$, where $q$, at this point, is an arbitrary parameter.

In order to allow the $p^+$ to decay, we shall endow the current
with an internal degree of freedom. For this purpose we shall promote 
$q$ to a self-adjoint operator $\hat q (\tau)$~\cite{DW}-\cite{BD} acting on 
a 2-dimensional Hilbert space associated with proton $|p \rangle$
and neutron $|n \rangle$ states. 
They will be assumed to be energy eigenstates of the proper 
free Hamiltonian $\hat H$ of the proton/neutron system:
$
\hat H |p  \rangle = m_{p } |p  \rangle\; ,
$
$
\hat H |n \rangle = m_n |n \rangle\; ,
$
where $m_p$ and $m_{n }$ are the $p^+$ and $n^0$ masses, 
respectively. In this context,  $|p  \rangle$ and 
$|n \rangle$ will be seen as unexcited and excited states of the nucleon, 
respectively. Further we will define the effective Fermi constant as
$G_F \equiv |\langle p | \hat q(0) | n \rangle |$, where
$ 
\hat q(\tau )\equiv e^{i\hat H \tau} \hat q(0) e^{-i\hat H \tau}\; 
$ 
and $\tau$ is the $p^+$ proper time.

In the inertial frame, the fermionic fields describing
the leptons in (i) can be written as 
\begin{equation}
\hat \Psi(t,z)= \sum_{\sigma = \pm } \int_{-\infty}^{+\infty} dk
\left( \hat a_{k \sigma} \psi^{(+\omega)}_{k \sigma} 
     + \hat c^\dagger_{k \sigma} \psi^{(-\omega)}_{-k -\sigma}  
\right)\;,
\label{FF}
\end{equation}
where $\omega = \sqrt{m^2 + k^2} \ge m$, and
$m$, $k$ and $\sigma$ represent mass, momentum, and 
polarization quantum numbers, respectively.
In the Dirac representation~\cite{IZ}, the Minkowski modes,
i.e., the ones defined with respect to the inertial Killing field 
$\partial/\partial t$, are 
$  
\psi^{(\pm \omega)}_{k \sigma} (t,z) \equiv 
\lambda^{(\pm \omega)}_{k \sigma}
{e^{i(\mp \omega t + kz)}}/{\sqrt{2\pi}} 
$
with
\begin{equation}
\lambda^{(\pm \omega)}_{k +} =
\left(
\begin{array}{c}
\pm \sqrt{(\omega \pm m)/2\omega} \\
0\\
k/\sqrt{2\omega(\omega \pm m)}\\
0
\end{array}
\right) \;\; 
\label{NM1}
\end{equation}
and
\begin{equation}
\lambda^{(\pm \omega)}_{k -}  = 
\left(
\begin{array}{c}
0\\
\pm \sqrt{(\omega \pm m)/2\omega} \\
0\\
-k/\sqrt{2\omega(\omega \pm m)}
\end{array}
\right) \;\; .
\label{NM2}
\end{equation}
Then the annihilation $ \hat a_{k \sigma} $, $ \hat c_{k \sigma} $ and creation 
$ \hat a^\dagger_{k \sigma} $, $ \hat c^\dagger_{k \sigma} $ operators satisfy
$
\{\hat a_{k \sigma},\hat a^\dagger_{k' \sigma'}\}=
$
$
\{\hat c_{k \sigma},\hat c^\dagger_{k' \sigma'}\}=
$
$
\delta(k-k') \; \delta_{\sigma \sigma'} 
$
and
$
\{\hat a_{k \sigma},\hat a_{k' \sigma'}\}=
$
$
\{\hat c_{k \sigma},\hat c_{k' \sigma'}\}=
$
$
\{\hat a_{k \sigma},\hat c_{k' \sigma'}\}=
$
$
\{\hat a_{k \sigma},\hat c^\dagger_{k' \sigma'}\}=
0. 
$

Let us assume that the electron and neutrino  
fields are coupled to the nucleon current according to 
the Fermi-like action
\begin{equation}
\hat S_I = \int d^2x \sqrt{-g} \hat j_\mu 
           (\hat{\bar \Psi}_\nu \gamma^\mu \hat \Psi_e +
            \hat{\bar \Psi}_e \gamma^\mu \hat \Psi_\nu ) \; .
\label{S}
\end{equation}
(The choice of other interaction actions   
would not change conceptually our final conclusions.)

\begin{figure}
\begin{center}
\mbox{\epsfig{file=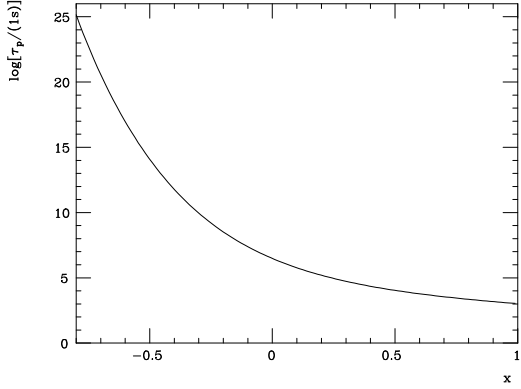,width=0.36\textwidth,angle=90}}
\end{center}
\vskip -1 cm
\caption{ The $p^+$ mean  proper lifetime is plotted 
as a function of $x \equiv \log_{10} (a/1\; {\rm MeV} )$, where
$ m_e=0.511 \;{\rm MeV}$ and $\Delta m = 1.29\; {\rm MeV}$.
($1 \; {\rm MeV} \approx 4.6 \times 10^{31} {\rm cm/s}^2$.)
Note that $\tau_p \to +\infty$ for inertial $p^+$'s.
}
\label{proton}
\end{figure}

The $p^+$ proper decay rate is written, thus, as
$$
{\Gamma}^{p  \to n}_{\rm (i)}= 
\frac{1}{T} 
\sum_{\sigma_e, \sigma_\nu=\pm} 
\int_{-\infty}^{+\infty} d k_e \int_{-\infty}^{+\infty} d k_\nu \;
| {\cal A}_{{\rm (i)}}^{p \to n} |^2 \;
$$
where  
$
{\cal A}^{p  \to n}_{\rm (i)} =
\; \langle  n \vert \otimes \langle e^+_{k_e \sigma_e} , 
\nu_{k_\nu \sigma_\nu} \vert \;
\hat S_I \;
\vert 0 \rangle \otimes \vert p  \rangle \; 
$
is the decay amplitude, at the tree level, and $T$ 
is the $p^+$ total {\em proper} time. 
Eventually, we obtain 
\begin{eqnarray}
& &
{\Gamma}^{p  \to n}_{\rm (i)}
 = 
\frac{G_F^2 \tilde m_e a }{2 \pi^{3/2} e^{\pi \widetilde{\Delta m}}}
\nonumber\\
& &
\times
G_{1\;3}^{3\;0} 
\left( \tilde m_e^2 \left|
\begin{array}{l}
\;\;\;1\\ 
-{1}/{2}\;,\;{1}/{2}+i \widetilde{\Delta m}\;,\;{1}/{2}-i \widetilde{\Delta m}
\end{array}
\right.
\right) ,
\label{RIF}
\end{eqnarray}
where 
$ \Delta m \equiv m_n - m_{p }$, $ \widetilde{\Delta m} \equiv \Delta m /a$,
$\tilde m_e \equiv m_e/a$, and we have assumed $m_\nu = 0$.
The value of the effective Fermi constant $G_F$
is fixed from phenomenology. 
By making $\Delta m \to - \Delta m$ and $a \to 0 $ in Eq.~(\ref{RIF}),
we  obtain that the mean proper lifetime of {\em inertial}
$n^0$'s due to $\beta$-decay,  
$$
{\rm (ii)} \;\;\; n^0 \to p^+ + e^- + \nu_e ,
$$ 
is 
$
\tau ^{n \to p }_{\rm (ii)}  
=
1/ {\Gamma}^{n \to p }_{\rm (ii)} 
=
\pi/( 2 G_F^2 \sqrt{{\Delta m}^2 -m_e^2 \; }\; ) .
$
Now let us assume that the $n^0$ mean lifetime 
in 2 dimensions is, e.g., $887~{\rm s}$. 
In this case, we obtain $G_F = 9.92 \times 10^{-13}$. 
Note that $G_F \ll 1$, which corroborates our perturbative approach. 
Now we are able to plot in Fig.~(\ref{proton}) the $p^+$  mean proper lifetime 
$
\tau ^{p \to n }_{\rm (i)} = 1/{\Gamma}^{p \to n }_{\rm (i)}
$ 
[see Eq.~(\ref{RIF})]
as a function of $a$. 
(The necessary energy to allow $p^+$'s to decay is provided
by the external accelerating agent.) 

Now let us describe the $p^+$ decay from the point of view of coaccelerated
observers according to which the $p^+$ is immersed in a FDU thermal 
bath at a temperature $T_{FDU}=a/2 \pi$. 
According to them, process (i) is  forbidden from energy 
conservation  (since the $p^+$ is static) but the following ones 
$$ 
{\rm (iii)}\, p^+ \, e^- \stackrel{a}{\to} n^0 \, \nu_e\, , 
{\rm (iv)}\, p^+ \, \bar \nu_e \stackrel{a}{\to} n^0 \, e^+ \, , 
{\rm (v)}\, p^+ \, e^- \bar \nu_e \stackrel{a}{\to} n^0  
$$ 
become  allowed since the $p^+$ can interact with the leptons of the thermal 
bath. By comparing process (i) against processes (iii)-(v), we can see how 
different are the descriptions given by the inertial and accelerated 
observers. 

The suitable coordinates to analyze the $p^+$ 
decay according to uniformly accelerated observers
are the Rindler ones $(v,u)$.  
They are related with the usual Cartesian coordinates  
by  $ t = u \sinh v \;, \;\; z = u \cosh v $,
where $0<u<+\infty$ and $-\infty<v<+\infty$.
In these coordinates,
the line element of  Minkowski spacetime at the Rindler wedge
($x > |t|$) is  
$
ds^2 = u^2 dv^2 - du^2 \; 
$
and the worldline of a $p^+$ with proper acceleration $a$ is 
$u = a^{-1} = const$.

According to uniformly accelerated observers, the fermionic field
is expanded  as~\cite{SMG}
\begin{equation}
\hat \Psi(v,u)= \sum_{\sigma = \pm } \int_{0}^{+\infty} d\bar \omega
\left( \hat b_{\bar \omega \sigma} \chi_{\bar \omega \sigma}
       + \hat d^\dagger_{\bar \omega \sigma} \chi_{-\bar \omega -\sigma} 
\right) ,
\label{PSA}
\end{equation}
where we recall that {\em Rindler} frequencies $\bar \omega$ may assume  
arbitrary positive real values since they do not obey any dispertion relation.
Here,
$\chi_{\bar \omega \sigma} (v,u) 
\equiv 
C_{\bar \omega} \xi_{\bar \omega \sigma} e^{-i\bar \omega v/a}$
where
$ 
C_{\bar \omega} \equiv \sqrt{[m \cosh (\pi \bar \omega/a)]/[2 \pi^2 a]} 
$
and
\begin{equation}
\xi_{\bar \omega +} =
\left(
\begin{array}{c}
K_{i \bar \omega/a + 1/2}(mu) +
i K_{i \bar \omega/a - 1/2}(mu)\\
0\\
-K_{i \bar \omega/a + 1/2}(mu) +
i K_{i \bar \omega/a - 1/2}(mu)\\
0
\end{array}
\right) ,
\label{RNM1}
\end{equation}
\begin{equation}
\xi_{\bar \omega -} =
\left(
\begin{array}{c}
0\\
K_{i \bar \omega/a + 1/2}(mu) +
i K_{i \bar \omega/a - 1/2}(mu)\\
0\\
K_{i \bar \omega/a + 1/2}(mu) -
i K_{i \bar \omega/a - 1/2}(mu)
\end{array}
\right) 
\label{RNM2}
\end{equation}
are positive and negative frequency {\em Rindler} modes, i.e., the ones defined 
with respect to the boost Killing field $a \partial / \partial v$. 
They are orthonormalized such that the annihilation 
$b_{\bar \omega \sigma}$, $d_{\bar \omega \sigma}$ and creation 
$b^\dagger_{\bar \omega \sigma}$, $d^\dagger_{\bar \omega \sigma}$
operators satisfy
$
\{\hat b_{\bar \omega \sigma},\hat b^\dagger_{\bar \omega' \sigma'}\}=
$
$
\{\hat d_{\bar \omega \sigma},\hat d^\dagger_{\bar \omega' \sigma'}\}=
$
$
\delta(\bar \omega-\bar \omega') \; \delta_{\sigma \sigma'} 
$
and also
$
\{\hat b_{\bar \omega \sigma},\hat b_{\bar \omega' \sigma'}\}=
$
$
\{\hat d_{\bar \omega \sigma},\hat d_{\bar \omega' \sigma'}\}=
$
$
\{\hat b_{\bar \omega \sigma},\hat d_{\bar \omega' \sigma'}\}=
$
$
\{\hat b_{\bar \omega \sigma},\hat d^\dagger_{\bar \omega' \sigma'}\}=
0.
$

The transition rates associated with processes (iii)-(v) are given by
\begin{eqnarray}
\Gamma^{p \to n}_{\rm (iii)} 
&=&
\frac{1}{T}
\sum_{\sigma_{e^-},\sigma_{\nu} =\pm}
\int_0^{+\infty}  d\bar \omega_{e^-}  
\int_0^{+\infty}  d\bar \omega_\nu
|{\cal A}^{p  \to n}_{{\rm (iii)}}|^2\nonumber
\\
&\times& n_F(\bar \omega_{e^-}) [1-n_F(\bar \omega_\nu)]\;,
\nonumber
\\
\Gamma^{p \to n}_{\rm (iv)}
&=&
\frac{1}{T}
\sum_{\sigma_{e^+},\sigma_{\bar\nu} =\pm}
\int_0^{+\infty} d\bar \omega_{e^+}  
\int_0^{+\infty}  d\bar \omega_{\bar\nu}
|{\cal A}^{p  \to n}_{{\rm (iv)}}|^2\nonumber
\\
&\times& n_F(\bar \omega_{\bar\nu}) [1-n_F(\bar \omega_{e^+})]\;,
\nonumber
\\
\Gamma^{p \to n}_{\rm (v)}
&=&
\frac{1}{T}
\sum_{\sigma_{e^-},\sigma_{\bar\nu} =\pm}
\int_0^{+\infty}  d\bar \omega_{e^-}  
\int_0^{+\infty}  d\bar \omega_{\bar\nu}
|{\cal A}^{p  \to n}_{{\rm (v)}}|^2 
\nonumber
\\
&\times& n_F(\bar \omega_{e^-}) n_F(\bar \omega_{\bar\nu})\;,
\nonumber
\end{eqnarray}
where at the tree level
$$
{\cal A}^{p  \to n}_{\rm (iii)} = 
\; \langle  n \vert \otimes \langle \nu_{\bar \omega_\nu \sigma_\nu} \vert \;
\hat S_I \;
\vert e^-_{\bar \omega_{e^-} \sigma_{e^-}} \rangle \otimes \vert p \rangle\;,
$$
$$
{\cal A}^{p  \to n}_{\rm (iv)} = 
\; 
\langle  n \vert \otimes \langle e^+_{\bar \omega_{e^+}\sigma_{e^+}} \vert
\;
\hat S_I \;
\vert {\bar \nu}_{\bar \omega_{{\bar \nu}} \sigma_{{\bar \nu}}} \rangle 
\otimes \vert p  \rangle\;,
$$
$$
{\cal A}^{p  \to n}_{\rm (v)} = 
\; \langle  n \vert  \otimes \langle 0 \vert \;
\hat S_I \;
\vert e^-_{\bar \omega_{e^-} \sigma_{e^-}}
      {\bar \nu}_{\bar \omega_{{\bar \nu}} \sigma_{{\bar \nu}}} \rangle 
      \otimes \vert p  \rangle\;,
$$
and we recall that in the Rindler wedge the $\gamma^\mu$ in $\hat S_I$ 
[see Eq.~(\ref{S})] should be  replaced by  $\gamma^\mu_R$
(see Ref.~\cite{SMG}). 
Here
$
n_F(\bar \omega) \equiv 1/(1+e^{\bar \omega/T_{\rm FDU}})
$
is the fermionic {\em thermal factor} which appears because of
the presence of the FDU thermal bath. 

After some calculations, we obtain
\begin{eqnarray}
\Gamma_{\rm (iii)}^{p\to n}
& = &
A
\int_{\widetilde{\Delta m}}^{+\infty} d \tilde{\bar \omega}_{e^-}
\frac{ K_{i\tilde {\bar \omega}_{e^-} + 1/2} (\tilde m_e)
       K_{i\tilde {\bar \omega}_{e^-} - 1/2} (\tilde m_e)}
{\cosh[\pi (\tilde {\bar \omega}_{e^-} -\widetilde{\Delta m}) ] }
\label{R5}
\nonumber
\\
\Gamma_{{\rm (iv)}}^{p\to n}
& = &
A
\int_{0}^{+\infty} d \tilde{\bar \omega}_{e^+}
\frac{ K_{i\tilde {\bar \omega}_{e^+} + 1/2} (\tilde m_e)
       K_{i\tilde {\bar \omega}_{e^+} - 1/2} (\tilde m_e)}
{\cosh[\pi (\tilde {\bar \omega}_{e^+} + \widetilde{\Delta m}) ] }
\label{R6}
\nonumber
\\
\Gamma_{{\rm (v)}}^{p\to n}
& = &
A
\int_{0}^{\widetilde{\Delta m}} d\tilde {\bar \omega}_{e^-}
\frac{ K_{i\tilde {\bar \omega}_{e^-} + 1/2} (\tilde m_e)
       K_{i\tilde {\bar \omega}_{e^-} - 1/2} (\tilde m_e)}
{\cosh[\pi (\tilde {\bar \omega}_{e^-} -\widetilde{\Delta m}) ] }
\label{R7}
\nonumber
\end{eqnarray}
where $A \equiv (G_F^2 \tilde m_e a)/(\pi^2  e^{\pi \widetilde{\Delta m}}) $.  
A branching ratio analysis~\cite{MV} indicates that for small 
accelerations, where 
``few'' high-energy particles are available in the FDU thermal bath, 
process (v) dominates over  processes (iii) and (iv), while for high 
accelerations, processes (iii) and (iv) dominate over  process (v).

The  $p^+$ total proper decay rate is obtained by adding up all contributions: 
\begin{eqnarray}
\Gamma_{\rm tot}^{p\to n}
&=&
\Gamma_{{\rm (iii)}}^{p\to n} 
+\Gamma_{{\rm (iv)}}^{p\to n}
+\Gamma_{{\rm (v)}}^{p\to n}
\nonumber
\\
&=&
A
\int_{-\infty}^{+\infty} d\tilde {\bar \omega}
\frac{ 
      K_{i\tilde {\bar \omega} + 1/2} (\tilde m_e)
      K_{i\tilde {\bar \omega} - 1/2} (\tilde m_e)   
     }
     {\cosh[\pi (\tilde {\bar \omega} -\widetilde{\Delta m}) ] }
 \; .
\label{ATRP2}
\end{eqnarray}
Now, $\Gamma_{\rm (i)}^{p\to n}$ and $\Gamma_{\rm tot}^{p\to n}$
must coincide. Eq.~(\ref{ATRP2}) is difficult to solve analytically 
because the integral variable is in the function index.
(This can be seen as reflecting the essentially distinct
inertial and coaccelerated frame calculations.) 
Hence we solve  Eq.~(\ref{ATRP2}) numerically. Finally, by plotting  
$\tau_{\rm tot}=1/\Gamma_{\rm tot}^{p\to n}$ 
as a function of $a$, we {\em precisely} obtain 
Fig.~(\ref{proton})~\cite{error}.
We emphasize that we would not have obtained any 
agreement if we did not assume the FDU effect. 
 
The confusion about what the FDU effect means
have led to erroneous conclusions including the one that this effect would 
not exist~\cite{Betal}. {\em For instance},  a $p^+$  with 
proper acceleration $a={\rm const}$ in the Minkowski vacuum
does {\em not} have to behave as if it were static in a (usual) Minkowski
thermal bath at a temperature $T=a/2\pi$. 
(The FDU effect does not ensure any such coincidence.)
The FDU effect can be rigorously derived~\cite{S} from
the general Bisognano and Wichmann's theorem~\cite{BW} obtained independently
from axiomatic QFT (which is not even restricted to linear quantum fields).
Moreover the necessity of the FDU effect for the consistency of 
the (successfully tested) standard QFT in Minkowski spacetime  means that this
effect was {\em already} observed~\cite{comp}. We have illustrated it
through the (potentially-important-to-astrophysics)
decay of accelerated $p^+$'s but other situations can be devised. 
Concerning  electromagnetic processes, e.g., the FDU thermal bath is 
crucial to reproduce the response of 
a uniformly accelerated $e^-$ to the Larmor radiation in the coaccelerated 
frame~\cite{HMS}. The same {\em must} be true if one takes into account the 
extra radiation induced by the $e^-$ recoil. 
There is no question about the existence of the FDU effect provided one 
accepts the validity of the results obtained with {\em standard} 
QFT in flat spacetime.

\begin{flushleft}
{\bf{\large Acknowledgements}}
\end{flushleft}
G.M. is deeply indebted to A.\ Higuchi, D.\ Sudarsky and
R.\ Wald for discussions on  QFT in noninertial frames
since long time. The authors would like to acknowledge particularly 
A.\ Higuchi for discussions  
at early stages of this work. G.M. and D.V. were supported 
by Conselho Nacional de Desenvolvimento Cient\'\i fico e 
Tecnol\'ogico (partially) and 
Funda\c c\~ao de Amparo \`a Pesquisa do Estado de S\~ao Paulo
(fully), respectively.


\begin{thebibliography}{99}
\bibitem{error}
Indeed, we get
$\Delta^{-1} \int_\Delta dx [(\tau_{\rm tot}-\tau_{\rm (i)})/\tau_{\rm tot}]^2
\sim 10^{-16}$, where $\Delta$ is the integration interval used in 
Fig.~(\ref{proton}).
\end{thebibliography}
\end{document}